# Inversion mechanism of uncertain orders and parameters for the non-commensurate and hyper fractional order chaotic systems via differential evolution ✩

GAO Fei[a,b,∗], FEI Feng-xia[a], XU Qian[a], DENG Yan-fang[a], QI Yi-bo[a]

[a]*Department of Mathematics, School of Science, Wuhan University of Technology, Luoshi Road 122, Wuhan, Hubei,430070, People's Republic of China*
[b]*Signal Processing Group, Department of Electronics and Telecommunications,Norwegian University of Science and Technology, N-7491 Trondheim, Norway*

## Abstract

In this paper, a novel uncertain fractional-orders and parameters' inversion mechanism via the differential evolution algorithms (DE) with a general mathematical model is proposed for non-commensurate and hyper fractional chaotic systems. The problems of fractional-order chaos' inversion estima-tion are converted into multiple modal non-negative objective functions' min-

✩The work was carried out during the tenure of the ERCIM Alain Bensoussan Fellow-ship Programme, which is supported by the Marie Curie Co-funding of Regional, National and International Programmes (COFUND) of the European Commission. The work is also supported by Supported by Scientific Research Foundation for Returned Scholars from Ministry of Education of China (No. 20111j0032), the HOME Program No. 11044(Help Our Motherland through Elite Intellectual Resources from Overseas) funded by China Association for Science and Technology, the NSFC projects No. 10647141, No.60773210 of China, the Natural Science Foundation No.2009CBD213 of Hubei Province of China, the Fundamental Research Funds for the Central Universities of China, the self–determined and innovative research funds of WUT No. 2012–Ia–035, 2012-Ia-041, 2010–Ia–004, The National Soft Science Research Program 2009GXS1D012 of China, the National Science Foundation for Post–doctoral Scientists of China No. 20080431004.
∗Corresponding author
*Email addresses:* hgaofei@gmail.com (GAO Fei ), 1092285218@qq.com (FEI Feng-xia), 121214020@qq.com (XU Qian), 986627833@qq.com (DENG Yan-fang), manshengyibo@163.com (QI Yi-bo)
*URL:* http://feigao.weebly.com (GAO Fei )

*Preprint submitted to Elsevier*                                                                *September 7, 2012*

abstractimization, which takes fractional-orders and parameters as its particular independent variables. And the objective is to find optimal combinations of fractional-orders and systematic parameters by DE in the predefined intervals for fractional order chaotic systems such that the objective function is minimized. Simulations are done to estimate a series of non-commensurate and hyper fractional chaotic systems. The experiments' results show that the proposed inversion mechanism for fractional-order chaotic systems is a successful methods with the advantages of high precision and robustness.

## 1. Introduction

Recently, the applications of fractional differential equations, whose nonlinear dynamics are described by a powerful tool with the concept of fractional calculus[1–9], began to appeal to related scientists[10–29] in following areas, bifurcation, hyperchaos, proper and improper fractional-order chaos systems and chaos synchronization[8, 12–14, 16–18, 24–32].

For the fractional-order chaos systems, most of the control and synchronization methods are generalized conclusive but much more complicated approaches from the methods for normal and hyper chaos system. However, when considering the controlling and synchronization for the fractional-order chaos systems, there are some systematic parameters and fractional orders are unknown. It is difficult to identify the parameters in the fractional-order chaotic systems with unknown parameters[16, 17, 33, 34].

The process to get the exact values of uncertain orders and parameters for the fractional order chaotic systems is called system inversion mechanism. Hitherto, there have been some approaches in system inversion for fractional-order chaos systems. For instances, synchronization for fractional-order chaos[33] and fractional order complex networks[34]. However, the design of controller and the updating law of parameter identification is still a hard task with techniques and sensitivities depending on the considered systems. And the non-classical way via artificial intelligence methods, for examples, differential evolution[16] and particle swarm optimization[17]. In which only the commensurate fractional order chaos systems and simplest case with one unknown fractional order for fractional-order chaos systems are discussed. And there exist basic hypotheses in traditional non-Lyapunov estimation methods[10, 16, 17]. That is, the parameters and fractional orders are partially known or the known data series coincide with definite fractional equationas forms $f = (f_1, f_2, ..., f_n)$ of fractional chaotic differential equations except some uncertain parameters and fractional orders $\Theta = (\theta_1, \theta_2, ..., \theta_n, q_1, q_2, ..., q_n)$. In which, the non- commensurate cases with different fractional orders $\{q_i, i = 1, ..., n\}$ are not included in the above non-Lyapunov ideas or not fully discussed either.

However, to the best of our knowledge, few work in non-classical way has been done to the parameters and fractional orders' inversion estimation of non-commensurate and hyper fractional-order chaos systems[35] so far.

The objective of this work is to present a novel simple but effective inversion mechanism of uncertain fractional-orders and parameters based on differential evolution algorithms (DE) to estimate the non-commensurate and hyper fractional order chaotic systems in a non-Lyapunov way. And the illustrative inversion simulations in different chaos systems system are discussed respectively.

The rest is organized as follows. Section 2 give a simple review on non-Lyapunove parameters inversion mechanism for fractional-order chaos systems. In Section 3, a general mathematical model for fractional chaos parameters inversion mechanism in non-Lyapunov way are analyzed. In section 4, firstly the DE are introduced briefly. Then a novel methods with proposed united model based on DE is proposed to estimate the fractional chaos systems. And simulations are done to a series of different non-commensurate and hyper fractional order chaotic systems. Conclusions are summarized briefly in Section 5.

## 2. Non-Lyapunove parameters' inversion mechanism for fractional order and normal chaos systems

We consider the following fractional-order chaos system.

$$\begin{cases} {}_\alpha D_t^q y_1(t) = f_1\left(y_1(t), ..., y_n(t), Y_0(t), \theta\right); \\ {}_\alpha D_t^q y_2(t) = f_2\left(y_1(t), ..., y_n(t), Y_0(t), \theta\right); \\ \quad \cdots \quad \cdots \quad \cdots \\ {}_\alpha D_t^q y_n(t) = f_n\left(y_1(t), ..., y_n(t), Y_0(t), \theta\right). \\ L = (y_1, y_2, ..., y_n) \end{cases} \quad (1)$$

Let $L(t) = (y_1(t), y_2(t), ..., y_n(t))^T \in \Re^n$ denotes the state vector. $\theta =$

$(\theta_1, \theta_2, ..., \theta_n)^T$ denotes the original parameters. $q = (q_1, q_2, ..., q_n), (0 < q_i < 1, i = 1, 2, ..., n)$ is the fractional derivative orders.

Normally the function $f = (f_1, f_2, ..., f_n)$ is known. And the $\theta = (\theta_1, \theta_2, ..., \theta_n)$, $q = (q_1, q_2, ..., q_n)$ will be the parameters to be estimated. Then a correspondent system are constructed as following.

$$\begin{cases} {}_aD_t^q \tilde{y}_1(t) = f_1\left(\tilde{y}_1(t), ..., \tilde{y}_n(t), Y_0(t), \theta\right); \\ {}_aD_t^q \tilde{y}_2(t) = f_2\left(\tilde{y}_1(t), ..., \tilde{y}_n(t), Y_0(t), \theta\right); \\ \quad \cdots \quad \cdots \quad \cdots \\ {}_aD_t^q \tilde{y}_n(t) = f_n\left(\tilde{y}_1(t), ..., \tilde{y}_n(t), Y_0(t), \theta\right). \\ \tilde{L} = (\tilde{y}_1, \tilde{y}_2, ..., \tilde{y}_n) \end{cases} \quad (2)$$

where $\tilde{L}(t), \tilde{\theta}, \tilde{q}$ are the correspondent variables to those in equation (1), and function $f$ are the same. The two systems (1) (2) have the same initial condition $Y_0(t)$.

Then the objective is obtained as following,

$$(\theta, q)^* = \arg\min_{(\theta, q)} F = \arg\min_{(\theta, q)} \sum_t \left\| L(t) - \tilde{L}(t) \right\|_2 \quad (3)$$

When some the fractional chaotic differential equations $f = (f_1, f_2, ..., f_n)$ are unknown, the objective will be,

$$(f_1, f_2, ..., f_n)^* = \arg\min_{(f_1, f_2, ..., f_n)} F \quad (4)$$

Equation (4) is fractional-order chaos' inversion mechanism called reconstruction. In Reference [36] a novel method was proposed to reconstruct the unknown equations $(f_1, f_2, ..., f_n)$ based on an united mathematical model. However, for the united mathematical model[36], to be identified is only $(f_1, f_2, ..., f_n)$ instead of $q$. That is, the left part $_\alpha D_t^q \tilde{Y}(t)$ of the equation (2) are not included.

Therefore, to estimate the $q$ of the equation (2) with unknown systematic parameters $\theta$ is still a question to be solved for parameters and orders estimation of non-commensurate and hyper fractional-order chaos systems.

Now we take the non-commensurate fractional order Lü system (5)[22, 37] for instance.

$$\begin{cases} _0D_t^{q_1} x(t) = a(y(t) - x(t)); \\ _0D_t^{q_2} y(t) = -x(t)z(t) + cy(t); \\ _0D_t^{q_3} z(t) = x(t)y(t) - bz(t). \end{cases} \quad (5)$$

when $(a, b, c) = (36, 3, 20)$, $(q_1, q_2, q_3) = (0.985, 0.99, 0.98)$, initial point $(0.2, 0.5, 0.3)$, Lü system (5) is chaotic. When $(a, b, c), (q_1, q_2, q_3)$ is unknown, we have to estimate them.

Secondly the objective function is chosen as:

$$p = F(a, b, c, q_1, q_2, q_3) = \sum_{t=0 \cdot h}^{T \cdot h} \left\| Y(t) - \tilde{Y}(t) \right\|_2 \quad (6)$$

The objective function for system (5) is shown as Figure 1.

Then the problems of inversion for chaotic system are transformed into that of nonlinear function optimization (6). And the smaller $p$ is, the better combinations of parameter $(a, b, c, q_1, q_2, q_3)$ is. The independent variables of these functions are $\theta = (a, b, c, q_1, q_2, q_3)$.

For the normal chaos system, parameters inversion of chaotic system is the chief task to be resolved and of vital significance, especially for some new hyperchaotic systems recently proposed[38–46]. And these systems can be thought as special case of the above fractional order chaos systems. And series of estimation methods are proposed based on genetic algorithms(GA), chaotic ant swarm algorithms, particle swarm optimizations(PSO), quantum PSO, artificial bee colony algorithms etc[47–61].

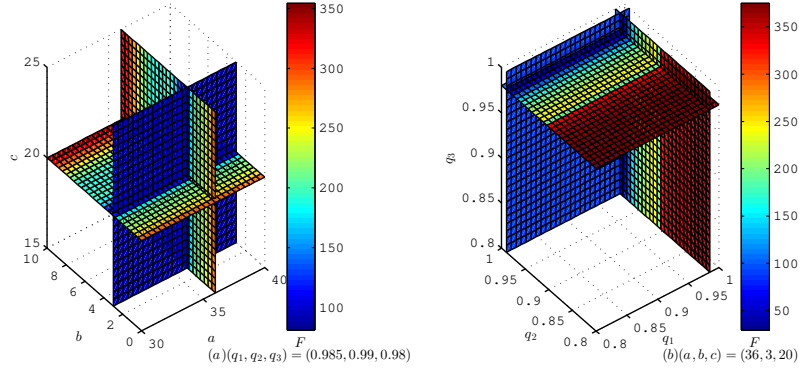

Figure 1: The objective function of fractional order Lü system

## 3. General mathematical model for fractional chaos inversion in non-Lyapunov way

In this section, a general mathematical model for fractional chaos parameters identification in non-Lyapunov way is proposed.

The general forms of fractional order chaos systems is (1). We consider that $(\tilde{f}_1, \tilde{f}_2, ..., \tilde{f}_n)$ is to be reconstructed and $(q_1, q_2, ..., q_n)$ is to be identified in (2).

To have simple forms, we take $\alpha = 0$. The continuous integro-differential operator[37, 62] is defined. And for the continuous function $f(t)$, the numerical solution method are used here[37, 62], which was obtained by the relationship derived from the G-L definition to resolve system.

Then the novel objective function (fitness) equation (7) in this paper come into being as below.

$$F = \sum_{t=0\cdot h}^{T\cdot h} \left\| \tilde{L} - L \right\|_2 \tag{7}$$

It should be noticed here that the independent variables in function (7) are not the parameters as in (6) but the special variables, for instance, as functions $(\tilde{f}_1, \tilde{f}_2, ..., \tilde{f}_n)$ and fractional orders $(q_1, q_2, ..., q_n)$. That is

$$\left( (\tilde{f}_1, \tilde{f}_2, ..., \tilde{f}_n), (\tilde{q}_1, \tilde{q}_2, ..., \tilde{q}_n) \right)^* = \underset{\left( (\tilde{f}_1, \tilde{f}_2, ..., \tilde{f}_n), (\tilde{q}_1, \tilde{q}_2, ..., \tilde{q}_n) \right)}{\arg\min} F \tag{8}$$

Equation (8) is the crucial turning point that changing from the parameters inversion into fractional equations and orders inversion, in other words, both fractional order estimation and fractional chaos systems' reconstruction.

Now, with the definitions of equations (7) and (8), a novel general mathematical model for fractional chaos parameters identification in non-Lyapunov way is coming into being.

It can be concluded that the parameters' estimation of fractional order chaos system[16, 17] is a special case of fractional order chaos reconstruction here as (8).

However, it should be emphasized here that, it is not easy to reconstruct the fractional order differential equations and identify the fractional orders together. And only the simplest case that with definite $q = q_1 = q_2 = ... = q_n$ are discussed[36].

And for a special cases of the general model (8) that $(q_1 \neq q_2 \neq ... \neq q_n)$ and systematic parameters $(\theta_1, \theta_2, ..., \theta_n) \in (f_1, f_2, ..., f_n)$ are unknown for non-commensurate and hyper fractional order chaos system, to our best of knowledge, no such inversion mechanisms have been done. Thus, it is necessary to resolve the following (9) in non-Lyapunov way.

$$(\tilde{q}, \tilde{\theta})^* = \arg\min_{(\tilde{q},\tilde{\theta})} F \tag{9}$$

## 4. A novel inversion mechanism based on differential evolution algorithm

The task of this section is to find a simple but effective inversion approach for unknown $q$ and systematic parameters in equation (9) of for non-commensurate and hyper fractional-order chaos based on Differential Evolution (DE) algorithm in non-Lyapunov way.

### 4.1. The Main Concept of Differential Evolution Algorithm

Differential Evolution (DE) algorithm grew out of Price's attempts to solve the Chebychev Polynomial fitting Problem that had been posed to him by Storn [63]. A breakthrough happened, when Ken came up with the idea of using vector differences for perturbing the vector population. Since this seminal idea, a lively discussion between Ken and Rainer and endless ruminations and computer simulations on both parts yielded many substantial improvements which make DE the versatile and robust tool it is today [63–66].

DE utilizes $M$ $n$–dimensional vectors, $X_i = (x_{i1}, \cdots, x_{in}) \in S, i = 1, \cdots, M$, as a population for each iteration, called a generation, of the algorithm. For each vector $X_i^{(G)} = (X_{i\,1}^{(G)}, X_{i\,2}^{(G)}, \cdots, X_{i\,n}^{(G)}), i = 1, 2, \cdots, M$, there are three main genetic operator acting.

For each individual, to apply the mutation operator, firstly random choose four mutually different individual in the current population $X_{r_1}^{(G)}, X_{r_2}^{(G)}, X_{r_3}^{(G)} (r_1 \neq r_2 \neq r_3 \neq i)$. Then combines it with the current best individual $X_{best}^{(G)}$ to get a perturbed vector $V = (V_1, V_2, \cdots, V_n)$ [63, 67] as below:

$$V = \begin{cases} X_{r_3}^{(G)} + 0.5(CF+1) \cdot \left(X_{r_1}^{(G)} + X_{r_2}^{(G)} - 2X_{r_3}^{(G)}\right), if\ rand(0,1) < 0.5 \\ X_{r_3}^{(G)} + CF \cdot \left(X_{r_1}^{(G)} - X_{r_2}^{(G)}\right), otherwise \end{cases} \tag{10}$$

where $CF > 0$ is a user-defined real parameter, called mutation constant, which controls the amplification of the difference between two individuals to avoid search stagnation.

Following the crossover phase, the crossover operator is applied on $X_i^{(G)}$. Then a trial vector $U = (U_1, U_2, \cdots, U_n)$ is generated by:

$$U_m = \begin{cases} V_m, & if\ (rand(0,1) < CR)\ or\ (m = k), \\ X_{i\,m}^{(G)}, & if\ (rand(0,1) \geq CR)\ and\ (m \neq k). \end{cases} \tag{11}$$

in the current population[63], where $m = 1, 2, \cdots, n$, the index $k \in \{1, 2, \cdots, n\}$ is randomly chosen, $CR$ is a user-defined crossover constant[63, 67] in the range $[0, 1]$. In other words, the trial vector consists of some of the components of the mutant vector, and at least one of the components of a randomly selected individual of the population.

Then it comes to the replacement phase. To maintain the population size, we have to compare the fitness of $U$ and $X_i^{(G)}$, then choose the better:

$$X_i^{(G+1)} = \begin{cases} U, if\ F(U) < F(X_i^{(G)}), \\ X_i^{(G)}, otherwise. \end{cases} \tag{12}$$

*4.2. A novel unknown parameters and fractional orders inversion mechanism*

Now we can propose a novel approach for hyper, proper and improper fractional chaos systems. The pseudo-code of the proposed reconstruction is given below.

---
**Algorithm 1** A novel inversion mechanism based on differential evolution algorithms
---
1: **Basic parameters' setting for DE**.
2: **Initialize** Generate the initial population.
3: **while** Termination condition is not satisfied **do**
4:     **Evaluation** Evaluate the fitness with Eq. (9) and remain the best individual.
5:     **Mutation** As in equation (10).
6:     **Crossover** As in equation (11).
7:     **Replacement** As in equation (12).
8:     **Boundary constraints** For each $x_{ik} \in X_i, k = 1, 2, ..., D$, if $x_{i1}$ is beyond the boundary, it is replaced by a random number in the boundary.
9: **end while**
10: **Output** Global optimum $x_{Best}$
---

*4.3. Non-commensurate and hyper fractional order chaos systems*

To test the Algorithm 1, some different well known and widely used non-commensurate and hyper fractional order chaos systemsare choose as following.

Example. 1. Fractional Lorénz system[13, 28, 68] for instance, which is generalized from the first canonical chaotic attractor found in 1963, Lorénz system[69].

$$\begin{cases} {}_\alpha D_t^{q_1} x = \sigma \cdot (y - x); \\ {}_\alpha D_t^{q_2} y = \gamma \cdot x - x \cdot z - y; \\ {}_\alpha D_t^{q_3} z = x \cdot y - b \cdot z. \\ L = (x, y, z) \end{cases} \quad (13)$$

where $q_1, q_2, q_3$ are the fractional orders. When $(q_1, q_2, q_3) = (0.985, 0.99, 0.99)$, $\sigma = 10, \gamma = 28, b = 8/3$, $\alpha = 0$, intimal point $(0.1, 0.1, 0.1)$, it is the non-commensurate chaotic system[68]. The objective function for system (13) is shown as following.

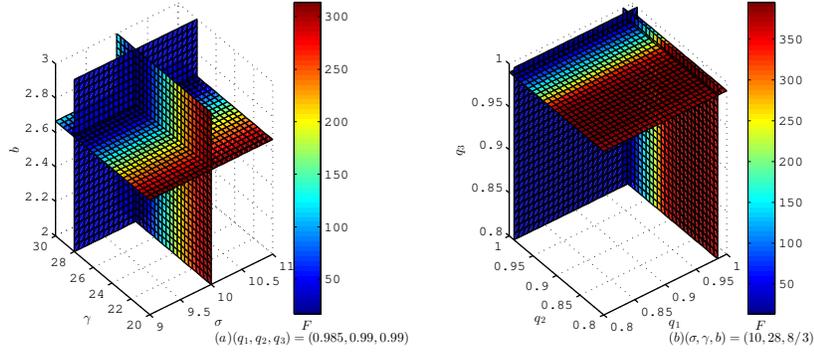

Figure 2: The objective function of fractional order Lorenz system

Example. 2. Fractional order Arneodo's System (14)[37, 70].

$$\begin{cases} {}_0 D_t^{q_1} x(t) = y(t); \\ {}_0 D_t^{q_2} y(t) = z(t); \\ {}_0 D_t^{q_3} z(t) = -\beta_1 x(t) - \beta_2 y(t) - \beta_3 z(t) + \beta_4 x^3(t). \end{cases} \quad (14)$$

when $(\beta_1, \beta_2, \beta_3, \beta_4) = (-5.5, 3.5, 0.8, -1.0)$, $(q_1, q_2, q_3) = (0.97, 0.97, 0.96)$, initial point $(-0.2, 0.5, 0.2)$, Arneodo's System (14) is chaotic.

Example. 3. Fractional order Duffing's system (15)[37].

$$\begin{cases} {}_0D_t^{q_1}x(t) = y(t); \\ {}_0D_t^{q_2}y(t) = x(t) - x^3(t) - \alpha y(t) + \delta \cos(\omega t). \end{cases} \quad (15)$$

when $(a, b, c) = (0.15, 0.3, 1)$, $(q_1, q_2) = (0.9, 1)$, initial point $(0.21, 0.31)$, Duffings system (15) is chaotic.

Example. 4. Fractional order Genesio-Tesi's System (16)[37, 71].

$$\begin{cases} {}_0D_t^{q_1}x(t) = y(t); \\ {}_0D_t^{q_2}y(t) = z(t); \\ {}_0D_t^{q_3}z(t) = -\beta_1 x(t) - \beta_2 y(t) - \beta_3 z(t) + \beta_4 x^2(t). \end{cases} \quad (16)$$

when $(\beta_1, \beta_2, \beta_3, \beta_4) = (1.1, 1.1, 0.45, 1.0)$, $(q_1, q_2, q_3) = (1, 1, 0.95)$, initial point $(-0.1, 0.5, 0.2)$, Genesio-Tesi's System (16) is chaotic.

Example. 5. Fractional order financial System (17)[37, 72] with the exact form of the differential equation ${}_0D_t^{q_3}z = f_3(x, y, z)$ are unknown.

$$\begin{cases} {}_0D_t^{q_1}x(t) = z(t) + x(t)(y(t) - a); \\ {}_0D_t^{q_2}y(t) = 1 - by(t) - x^2(t); \\ {}_0D_t^{q_3}z(t) = -x(t) - cz(t). \end{cases} \quad (17)$$

when $(a, b, c) = (1, 0.1, 1)$, $(q_1, q_2, q_3) = (1, 0.95, 0.99)$, initial point $(2, -1, 1)$, financial System (17) is chaotic.

Example. 6. Fractional order Lü system (5).

Example. 7. Improper fractional order Chen system (18)[37, 73, 74].

$$\begin{cases} {}_0D_t^{q_1}x(t) = a(y(t) - x(t)); \\ {}_0D_t^{q_2}y(t) = (d)x(t) - x(t)z(t) + cy(t); \\ {}_0D_t^{q_3}z(t) = x(t)y(t) - bz(t). \end{cases} \quad (18)$$

And when when $(a, b, c, d) = (35, 3, 28, -7)$, $(q_1, q_2, q_3) = (1, 1.24, 1.24)$, initial point $(3.123, 1.145, 2.453)$, Chen system (18) is an improper chaotic system[18].

Example. 8. Fractional order Rössler System (19)[37, 75].

$$\begin{cases} {}_0D_t^{q_1} x(t) = -(y(t) + z(t)); \\ {}_0D_t^{q_2} y(t) = x(t) + ay(t); \\ {}_0D_t^{q_3} z(t) = b + z(t)(x(t) - c). \end{cases} \quad (19)$$

when $(a, b, c) = (0.5, 0.2, 10)$, $(q_1, q_2, q_3) = (0.9, 0.85, 0.95)$, initial point $(0.5, 1.5, 0.1)$, Rössler System (19) is chaotic.

Example. 9. Fractional order Chuas oscillator (20)[76].

$$\begin{cases} {}_0D_t^{q_1} x(t) = \alpha(y(t) - x(t) + \zeta x(t) - W(w) x(t)); \\ {}_0D_t^{q_2} y(t) = x(t) - y(t) + z(t); \\ {}_0D_t^{q_3} z(t) = -\beta y(t) - \gamma z(t); \\ {}_0D_t^{q_4} w(t) = x(t); \end{cases} \quad (20)$$

where

$$W(w) = \begin{cases} a : |w| < 1; \\ b : |w| > 1. \end{cases}$$

when $(\alpha, \beta, \gamma, \zeta, a, b) = (10, 13, 0.1, 1.5, 0.3, 0.8)$, $(q_1, q_2, q_3, q_4) = (0.97, 0.97, 0.97, 0.97)$, initial point $(0.8, 0.05, 0.007, 0.6)$, Chua's oscillator (20) is chaotic.

Example. 10. Hyper fractional order Lorénz System (21)[77].

$$\begin{cases} {}_0D_t^{q_1} x(t) = a(y(t) - x(t)) + w(t); \\ {}_0D_t^{q_2} y(t) = cx(t) - x(t) z(t) - y(t); \\ {}_0D_t^{q_3} z(t) = x(t) y(t) - bz(t); \\ {}_0D_t^{q_4} w(t) = -y(t) z(t) + \gamma w(t); \end{cases} \quad (21)$$

when $(a, b, c, d) = (10, 8/3, 28, -1)$, $(q_1, q_2, q_3, q_4) = (0.96, 0.96, 0.96, 0.96)$, initial point $(0.5, 0.6, 1, 2)$, Hyper fractional order Lorénz System (21) is chaotic.

Example. 11. Hyper fractional order Lü System (22)[78].

$$\begin{cases} {}_0D_t^{q_1}x(t) = a(y(t) - x(t)) + w(t); \\ {}_0D_t^{q_2}y(t) = -x(t)z(t) + cy(t); \\ {}_0D_t^{q_3}z(t) = x(t)y(t) - bz(t); \\ {}_0D_t^{q_4}w(t) = x(t)z(t) + dw(t); \end{cases} \quad (22)$$

when $(a, b, c, d) = (36, 3, 20, 1.3)$, $(q_1, q_2, q_3, q_4) = (0.98, 0.980.98, 0.98)$, initial point $(1, 1, 1, 1)$, Hyper fractional order Lü System (22) is chaotic.

Example. 12. Hyper fractional order Liu System (23)[79].

$$\begin{cases} {}_0D_t^{q_1}x(t) = -ax(t) + by(t)z(t) + z(t); \\ {}_0D_t^{q_2}y(t) = 2.5y(t) - x(t)z(t); \\ {}_0D_t^{q_3}z(t) = x(t)y(t) - cz(t) - 2w(t); \\ {}_0D_t^{q_4}w(t) = -d \cdot x(t). \end{cases} \quad (23)$$

when $(a, b, c, d) = (10, 1, 4, 0.25)$, $(q_1, q_2, q_3, q_4) = (0.9, 0.9, 0.9, 0.9)$, initial point $(2.4, 2.2, 0.8, 0)$, Hyper fractional order Liu System (23) is chaotic. The objective function for system (23) is shown as following.

Example. 13. Hyper fractional order Chen System (24)[80].

$$\begin{cases} {}_0D_t^{q_1}x(t) = -a(y(t) - x(t)) + w(t); \\ {}_0D_t^{q_2}y(t) = dx(t) - x(t)z(t) + cy(t); \\ {}_0D_t^{q_3}z(t) = x(t)y(t) - bz(t); \\ {}_0D_t^{q_4}w(t) = y(t)z(t) + rw(t). \end{cases} \quad (24)$$

when $(a, b, c, d) = (35, 3, 12, 7, 0.5)$, $(q_1, q_2, q_3, q_4) = (0.96, 0.96, 0.96, 0.96)$, initial point $(0.5, 0.6, 1, 2)$, Hyper fractional order Chen System (24) is chaotic.

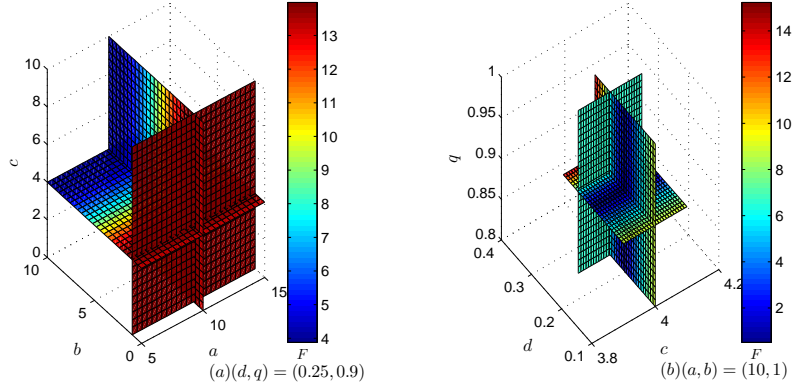

Figure 3: The objective function of Hyper fractional order Liu System system

Example. 14. Hyper fractional order Rössler System (25)[75].

$$\begin{cases} {}_0D_t^{q_1} x(t) = -(y(t) + z(t)); \\ {}_0D_t^{q_2} y(t) = x(t) + ay(t) + w(t); \\ {}_0D_t^{q_3} z(t) = x(t)z(t) + b; \\ {}_0D_t^{q_4} w(t) = -cz(t) + dw(t). \end{cases} \quad (25)$$

when $(a, b, c, d) = (0.32, 3, 0.5, 0.05)$, $(q_1, q_2, q_3, q_4) = (0.95, 0.950.95, 0.95)$, initial point $(-15.5, 9.3, -4, 18.6)$, Hyper fractional order Rössler System (25) is chaotic.

Example. 15. A four-wing fractional order system[81, 82] both incommensurate and hyper chaotic.

$$\begin{cases} D_t^{q_1} x_1 = ax_1 - x_2 x_3 + x_4, \\ D_t^{q_2} x_2 = -bx_2 + x_1 x_3, \\ D_t^{q_3} x_3 = x_1 x_2 - cx_3 + x_1 x_4, \\ D_t^{q_4} x_4 = -x_2, \end{cases} \quad (26)$$

when $(a, b, c) = (8, 40, 49)$, $(q_1, q_2, q_3, q_4) = (1, 0.95 0.9, 0.85)$, initial point $(1, -2, 3, 1)$[81], system (26) is chaotic.

## 4.4. Simulations

For systems to be identified, the parameters of the proposed method are set as following. The parameters of the simulations are fixed: the size of the population was set equal to $M = 40$, generation is set to 500, the default values $CF = 1$, $CR = 0.85$; The times of function evaluations are 20040. Table 1 give the detail setting for each system.

Table 2 shows the simulation results of above fractional order chaotic systems.

The following figures give a illustration how the self growing evolution process works by DE Algorithm 1. In which, Figures 4,5 ,6,7 8 , 9 , 10,11 show the simulation evolution results of above fractional order chaotic systems with optimization process of objective function's evolution and the parameters and orders uncertain of above fractional order chaotic systems.

From the simulations results of reconstruction above fractional order chaos system, it can be concluded that the proposed method is efficient. And from above figures, it can be concluded that the estimated systems are self growing under the genetic operations of the proposed methods.

To test the performance of the proposed method Algorithms 1 , some more simulations are done to the four-wing incommensurate hyper fractional order chaotic system (26) in following cases A,B,C,D. In these cases, each with only one condition is changed according to the original setting for system (26). And the simulation results are listed in Table 3.

Table 1: Detail parameters stetting for different systems

| F-O systems | Unknown[a] | Lower boundary | Upper boundary | Step | No. of samples |
|---|---|---|---|---|---|
| Lorénz | $(\sigma, \gamma, b, q_1, q_2, q_3)$ | $5, 20, 0.1, 0.1, 0.1, 0.1$ | $15, 30, 10, 1, 1, 1$ | $0.01$ | $100$ |
| Arneodo | $(\beta_1, \beta_2, \beta_3, \beta_4, q_1, q_2, q_3)$ | $-6, 2, 0.1, -1.5, 0.1, 0.1, 0.1$ | $-5, 5, 1, -0.5, 1, 1, 1$ | $0.005$ | $200$ |
| Duffing | $(a, b, c, q_1, q_2)$ | $0.1, 0.1, 0.1, 0.1, 0.5$ | $1, 1, 2, 1, 1.5$ | $0.0005$ | $500$ |
| Genesio-Tesi | $(\beta_1, \beta_2, \beta_3, \beta_4, q_1, q_2, q_3)$ | $1, 1, 0.1, 0.1, 0.5, 0.5, 0.1$ | $2, 2, 1, 1.5, 1.5, 1.5, 1$ | $0.005$ | $200$ |
| Financial | $(a, b, c, q_1, q_2, q_3)$ | $0.5, 0.01, 0.5, 0.5, 0.1, 0.1$ | $1.5, 1, 1.5, 1.5, 1, 1$ | $0.005$ | $200$ |
| Lü | $(a, b, c, q_1, q_2, q_3)$ | $30, 0.1, 15, 0.1, 0.1, 0.1$ | $40, 10, 25, 1, 1, 1$ | $0.01$ | $100$ |
| Improper Chen | $(a, b, c, d, q_1, q_2, q_3)$ | $30, 0.1, 20, -10, 0.5, 1, 1$ | $40, 10, 30, -0.1, 2, 2, 2$ | $0.01$ | $100$ |
| Rössler | $(a, b, c, q_1, q_2, q_3)$ | $0.1, 0.1, 5, 0.1, 0.1, 0.1$ | $1, 1, 15, 1, 1, 1$ | $0.01$ | $100$ |
| ChuaM | $(\alpha, \beta, \gamma, \zeta, a, b, q)$ | $5, 10, 0.1, 0.1, 0.3, 0.1, 0.1$ | $10, 20, 1, 2, 0.3, 1, 1$ | $0.01$ | $100$ |
| Hyper Lorénz | $(a, b, c, d, q)$ | $5, 0.1, 20, -2, 0.1$ | $15, 5, 30, -0.1, 1$ | $0.01$ | $100$ |
| Hyper Lü | $(a, b, c, d, q)$ | $30, 0.1, 15, 0.1, 0.1$ | $40, 5, 25, 5, 1$ | $0.005$ | $200$ |
| Hyper Liu | $(a, b, c, d, q)$ | $5, 0.5, 1, 0.1, 0.1$ | $15, 1.5, 10, 1, 1$ | $0.005$ | $100$ |
| Hyper Chen | $(a, b, c, d, q)$ | $30, 0.1, 10, 0.1, 0.1, 0.1$ | $40, 5, 20, 10, 1, 1$ | $0.005$ | $200$ |
| Hyper Rössler | $(a, b, c, d, q)$ | $0.1, 0.1, 0.1, 0.01, 0.1$ | $1, 5, 1, 1, 1$ | $0.005$ | $200$ |
| System (26) | $(a, b, c, q_1, q_2, q_3, q_4)$ | $5, 38, 45, 0.5, 0.5, 0.5, 0.5$ | $10, 45, 50, 1, 1, 1, 1$ | $0.005$ | $200$ |

[a] For the hyper fractional order chaos system, let $q = q_1 = q_2 = q_3 = q_4$.

Table 2: Simulation results for different fractional order chaos systems

| F-O system | StD | Mean | Min | Max | Success rate[a] |
|---:|---|---|---|---|---|
| Lorénz | 1.2271e-05 | 2.3801e-05 | 3.6823e-06 | 5.4275e-05 | 100% |
| Arneodo | 6.1944e-07 | 1.4863e-06 | 4.6606e-07 | 4.3293e-06 | 100% |
| Duffing | 2.4378e-10 | 2.4788e-10 | 2.867e-11 | 1.7728e-09 | 100% |
| Genesio-Tesi | 1.76e-06 | 3.7268e-06 | 6.6136e-07 | 1.0572e-05 | 100% |
| Financial | 2.3322e-07 | 4.5927e-07 | 1.1207e-07 | 1.1711e-06 | 100% |
| Lü | 1.8797e-05 | 2.9371e-05 | 5.2309e-06 | 1.0775e-04 | 99% |
| Improper Chen | 2.5763e-03 | 4.1358e-03 | 6.6422e-04 | 1.8223e-02 | 97% [b] |
| Rössler | 8.4114e-08 | 1.5306e-07 | 2.4063e-08 | 4.5141e-07 | 100% |
| ChuaM | 1.6998e-06 | 2.7689e-06 | 4.2085e-07 | 7.9197e-06 | 100% |
| Hyper Lorénz | 7.2263e-08 | 9.5764e-08 | 1.4287e-08 | 3.8946e-07 | 100% |
| Hyper Lü | 5.9079e-08 | 1.037e-07 | 2.2241e-08 | 2.9391e-07 | 100% |
| Hyper Liu | 1.0783e-09 | 1.6299e-09 | 1.3241e-010 | 6.2280e-09 | 100% |
| Hyper Chen | 1.1798e-05 | 2.0742e-05 | 6.1987e-06 | 6.0546e-05 | 100% |
| Hyper Rössler | 6.1201e-09 | 6.9673e-09 | 1.7082e-09 | 4.9026e-08 | 100% |
| System (26) | 2.9764e-03 | 5.7449e-03 | 1.7212e-03 | 1.6418-02 | 90% [b] |

[a] Success means the the solution is less than $1e-4$ in 100 independent simulations.

[b] Success means the the solution is less than $1e-2$ in 100 independent simulations.

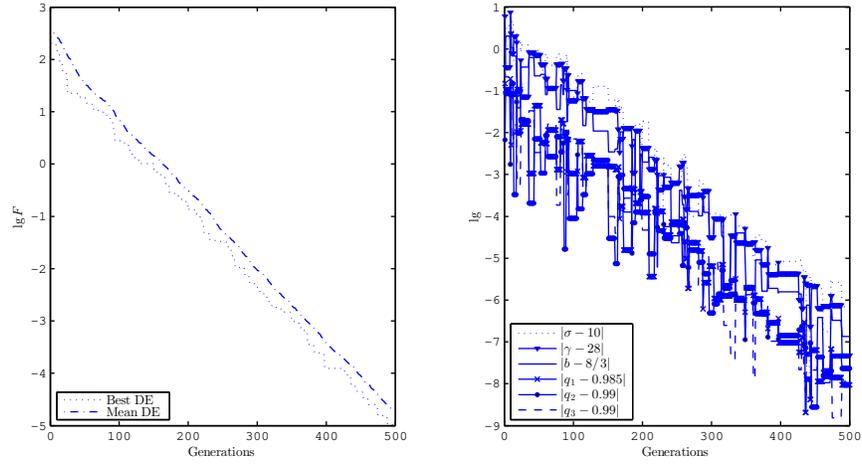

Figure 4: Evolution process for fractional order Lorenz system

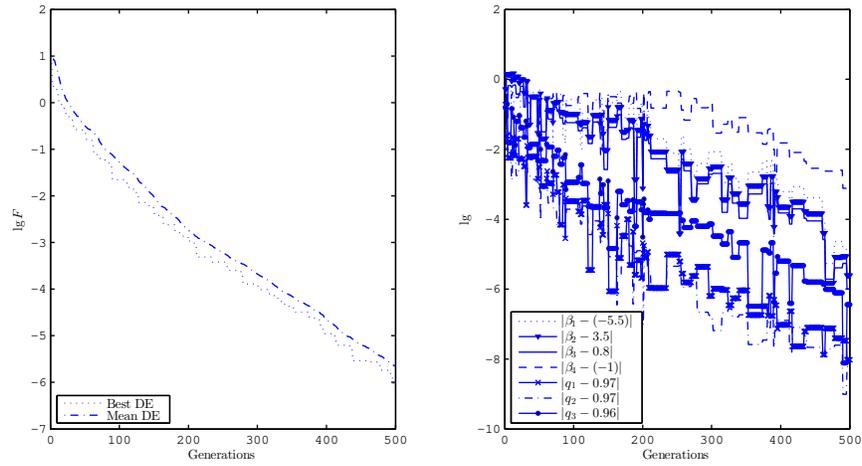

Figure 5: Evolution process for fractional order Arneodo system

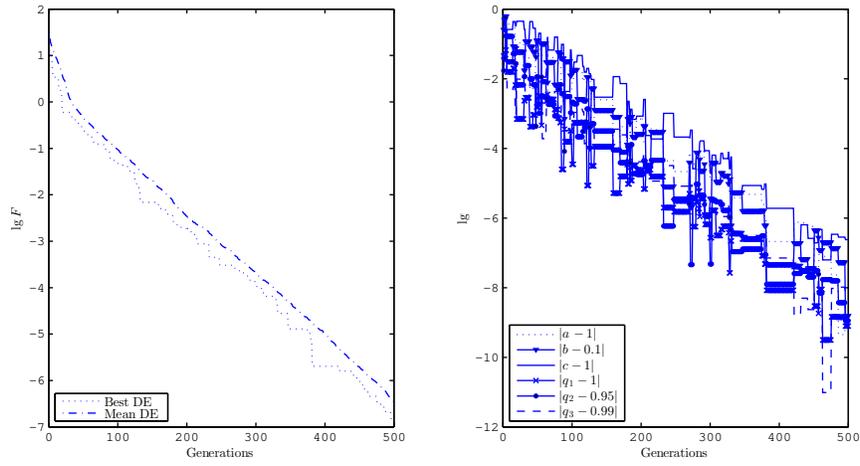

Figure 6: Evolution process for fractional order Financial system

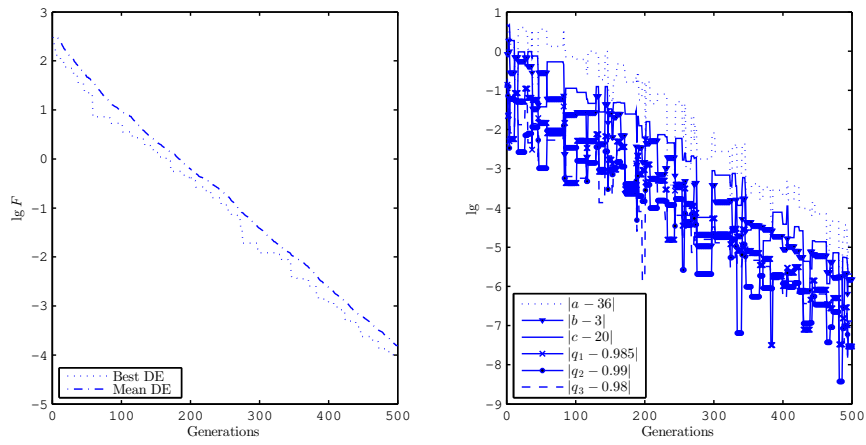

Figure 7: Evolution process for fractional order Lü system

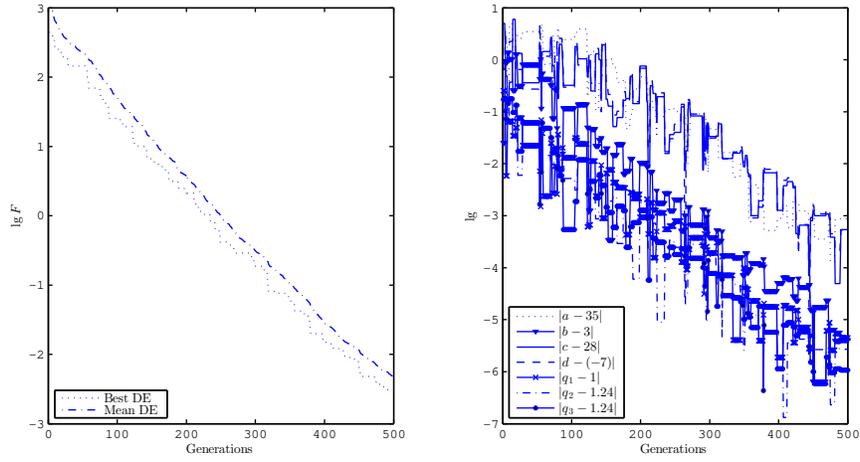

Figure 8: Evolution process for fractional order improper Chen system

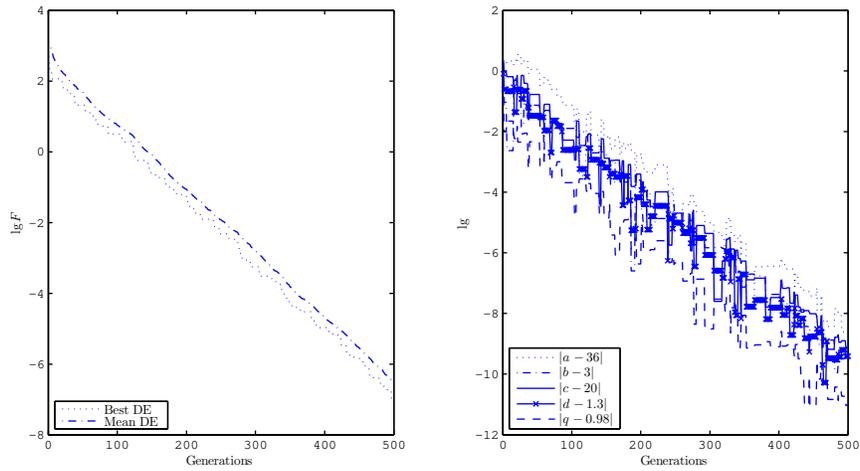

Figure 9: Evolution process for fractional order hyper Lü system

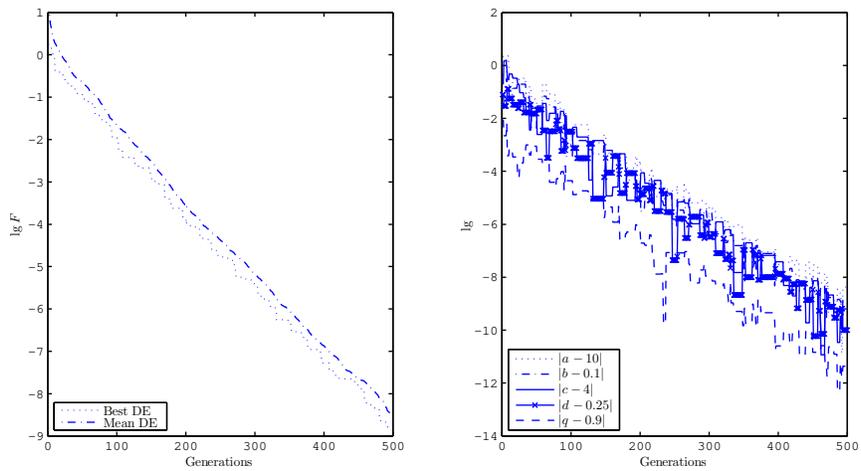

Figure 10: Evolution process for fractional order hyper Liu system

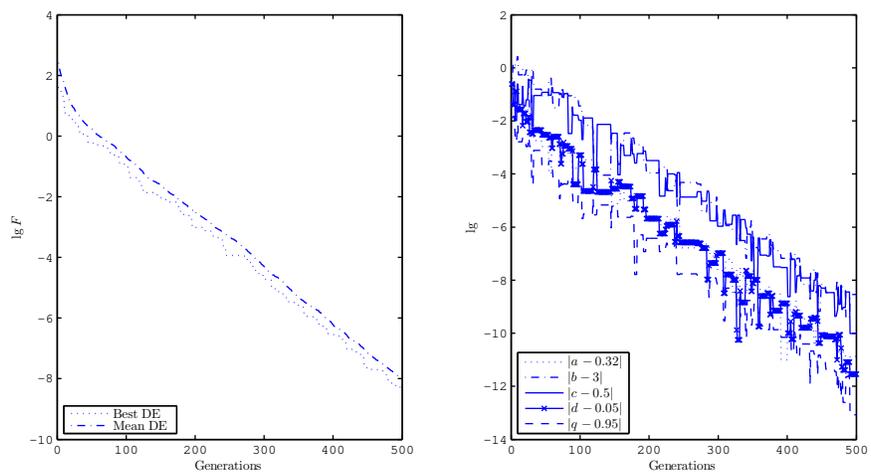

Figure 11: Evolution process for fractional order hyper Rössler system

- Case A. Enhancing the defined intervals of the unknown parameters and orders to $[0, 10] \times [30, 40] \times [40, 50] \times [0.1, 1] \times [0.1, 1] \times [0.1, 1] \times [0.1, 1]$.

- Case B. Minimizing the number of samples for computing system (26) from 200 to 100.

- Case C. Changing the iteration numbers of Algorithms 1 from 500 to 800.

- Case D. Changing the population size of Algorithms 1 from 40 to 80.

Table 3: Simulation results for system (26)

| system (26) | StD | Mean | Min | Max | Success rate[a] | NEOF[b] |
|---|---|---|---|---|---|---|
| Case A. | 3.6259e+02 | 3.5359e+02 | 4.3690e-03 | 7.2477e+02 | 20% | 20040 |
| Case B. | 5.2549e-05 | 1.1430e-04 | 3.8482e-05 | 2.7105e-04 | 100% | 20040 |
| Case C. | 5.8123e-06 | 1.0498e-05 | 1.7082e-06 | 3.0149e-05 | 100% | 32040 |
| Case D. | 3.3265e-03 | 9.6975e-03 | 2.9053e-03 | 2.1081e-02 | 64% | 40080 |

[a] Success means the the solution is less than $1e-2$ in 100 independent simulations.

[b] No. of evaluation for objective function

Figure 12 show the coresspondent simulation results for system (26).

From results of the Table 2, 3 and Figure 12, we can conclude that minimizing the number of samples for computing the system (26) as case B, enhancing the iteration numbers as case C, the population size of Algorithms 1 as case D, will make the Algorithms 1 much more efficient and achieve a much more higher precision. However if the defined intervals of the unknown parameters of system (26) are enhanced, then the results will go to the opposite way. That is the success rate is from 90% to 20% as case A.

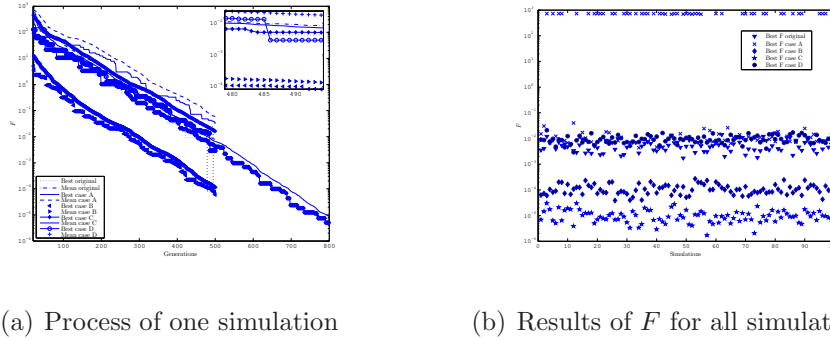

(a) Process of one simulation  (b) Results of $F$ for all simulations

Figure 12: Simulation results for system (26)

And considering the No. of evaluation for objective function, it is that minimizing the number of samples for computing the system (26) as case B is the best way to achieve higher efficiency and precision.

## 5. Conclusions

The inversion mechanism put consists of numerical optimization problem with unknown fractional order differential equations to identify the chaotic systems. Simulation results demonstrate the effectiveness and efficiency of the proposed methods with the mathematical model in Section 3. This is a novel Non–Lyaponov way for fractional order chaos' unknown parameters and orders.

The performance of the proposed method is sensitive to a series factors, such as the initial point for each fractional order chaos system, sample interval, number of points, and length of intervals for the unknown orders and parameters. Actually, these also lead to the candidate system divergent. And they are not predefined randomly. A good combination of these is not easy to get. Some mathematical formula to get a good combination not by so many simulations will be introduced in the future studies.

It should be noticed too many points for evaluating fractional order chaos system the individual represents are not worth. Because the most time consumption parts in the whole proposed method are to resolve the candidate systems. Some of these system are easy to solve. However when it comes with the some individuals with bad combinations of parameters and orders, the methods to resolve the fractional order chaos systems in Section 3 might not converge as shown in the simulations for system (26). Then the whole proposed method Algorithms 1 might get into endless loops. To avoid the endless loops, we introduce a forced strategy to assign all the NAN and infinite numbers in the output as zero. Because the objective function (9) to be optimized is bigger than 0, so this forced strategy for assignment is reasonable. To achieve a fine balance between the performance of the proposed methods and having enough sample data for credibility, we take the number of the points as $100 - 200$, according the existing simulations[47–61]. And the simulations in section 4 results show it is effective too.

Here we have to say that this work is only about the estimation of unknown parameters and orders with the objective function (9) for for non-commensurate and hyper fractional order chaos systems in non-Lyapunov way. It can be concluded that DE in Algorithms 1 can be change to other artificial intelligence methods easily. For the cases that some fractional order differential equations are unknown but with definite orders $q$ have been discussed in Reference [36].

In the future, we will do further researches for the cases that neither the fractional orders nor some fractional order equations are known. That is, the objective function is chosen as (8) in the novel mathematic model in Section 3. In another words, the objective will be changed into as equation (27).

$$(\tilde{q}, \tilde{f})^* = \arg\min_{(\tilde{q}, \tilde{f})} F \qquad (27)$$

In conclusion, it has to be stated that proposed Algorithms 1 for fractional order chaos systems' identification in a non-Lyapunov way is a promising direction.